\documentclass{kapproc} 

\RequirePackage{graphicx}%
\RequirePackage{epsf}%
\input{psfig.sty}

\upperandlowercase
\let\footnote\savefootnote
\let\footnotetext\savefootnotetext 
\let\footnoterule\savefootnoterule 
\kluwerbib 

\begin{document}


\articletitle{The Initial Mass Function 
in disc galaxies and in galaxy clusters: 
the chemo-photometric picture}

\chaptitlerunninghead{The IMF in disc galaxies and in galaxy clusters}



 \author{Laura Portinari}
 \affil{Tuorla Observatory, University of Turku, V\"ais\"alantie 20, FIN-21500
Pikki\"o, Finland}
 \email{lporti@utu.fi}





 \begin{abstract}
The observed brightness of the Tully-Fisher relation suggests a low stellar 
M/L ratio and a ``bottom-light'' IMF in disc galaxies, but
the corresponding efficiency of chemical enrichment tends to exceed 
the observational estimates. Either suitable tuning 
of the IMF slope and mass limits
or metal outflows from disc galaxies 
must then be invoked.

A standard Solar Neighbourhood IMF cannot explain the high metallicity 
of the hot intra-cluster medium: a different IMF must be at work
in clusters of galaxies.
Alternatively, if the IMF is universal and chemical enrichment
is everywhere as efficient as observed in clusters,
substantial loss of metals must occur 
from the Solar Neighbourhood 
and from disc galaxies in general; a "non-standard" scenario
challenging our
understanding of 
disc galaxy formation. 
 \end{abstract}

\section{The M$_*$/L ratio and the IMF in disc galaxies}
\label{sect:discIMF}

\begin{figure}
\sidebyside
{\centerline{\psfig{file=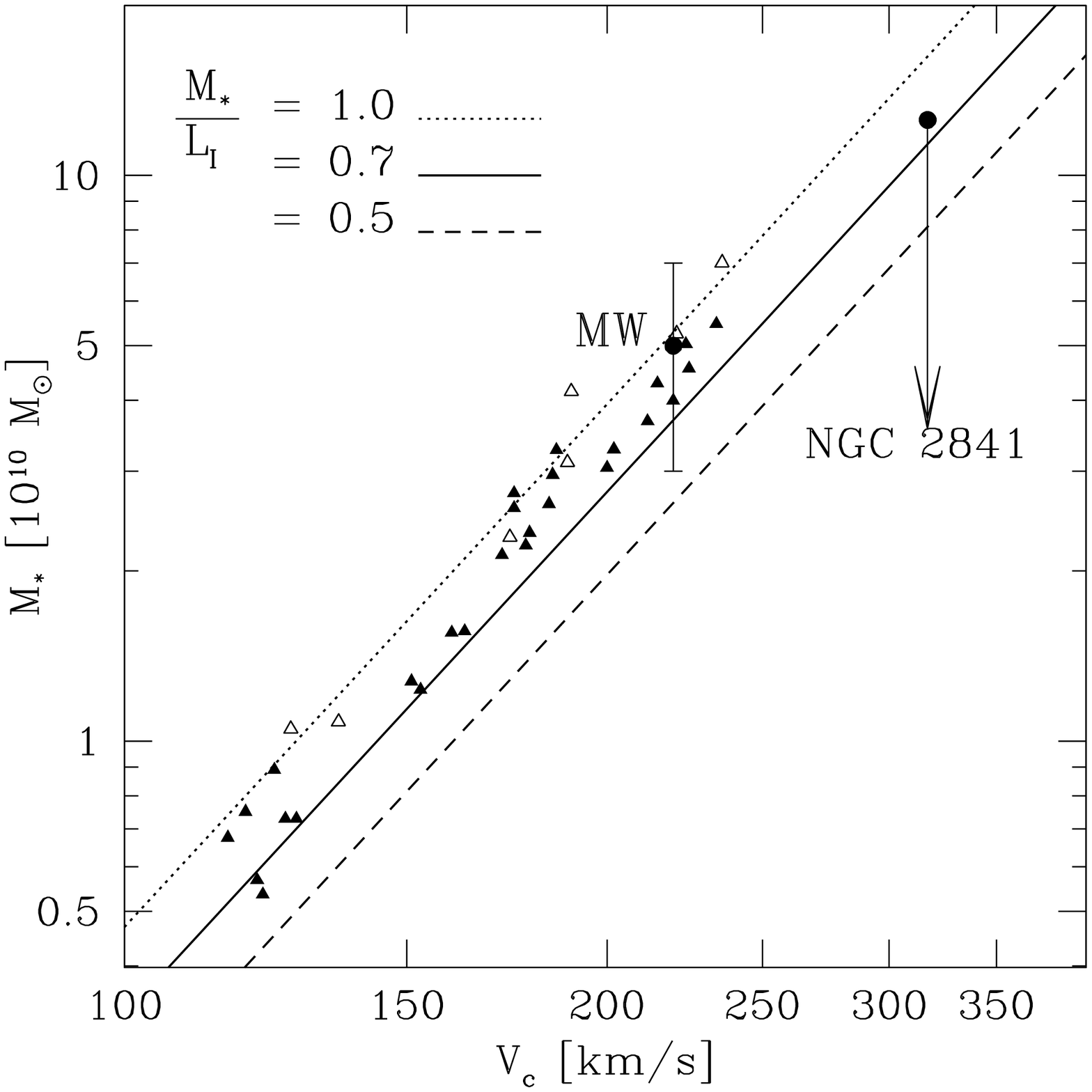,width=6truecm}}
\caption{
Observed Tully--Fisher relation for Sbc--Sc 
spirals (Dale et~al.\ 1999; $h$=0.7), 
assuming different M$_*$/L$_I$. 
{\em Triangles}: simulated galaxies;
{\em full dots}: 
Milky Way and NGC 2841.}
\label{fig:tully}}
{\centerline{\psfig{file=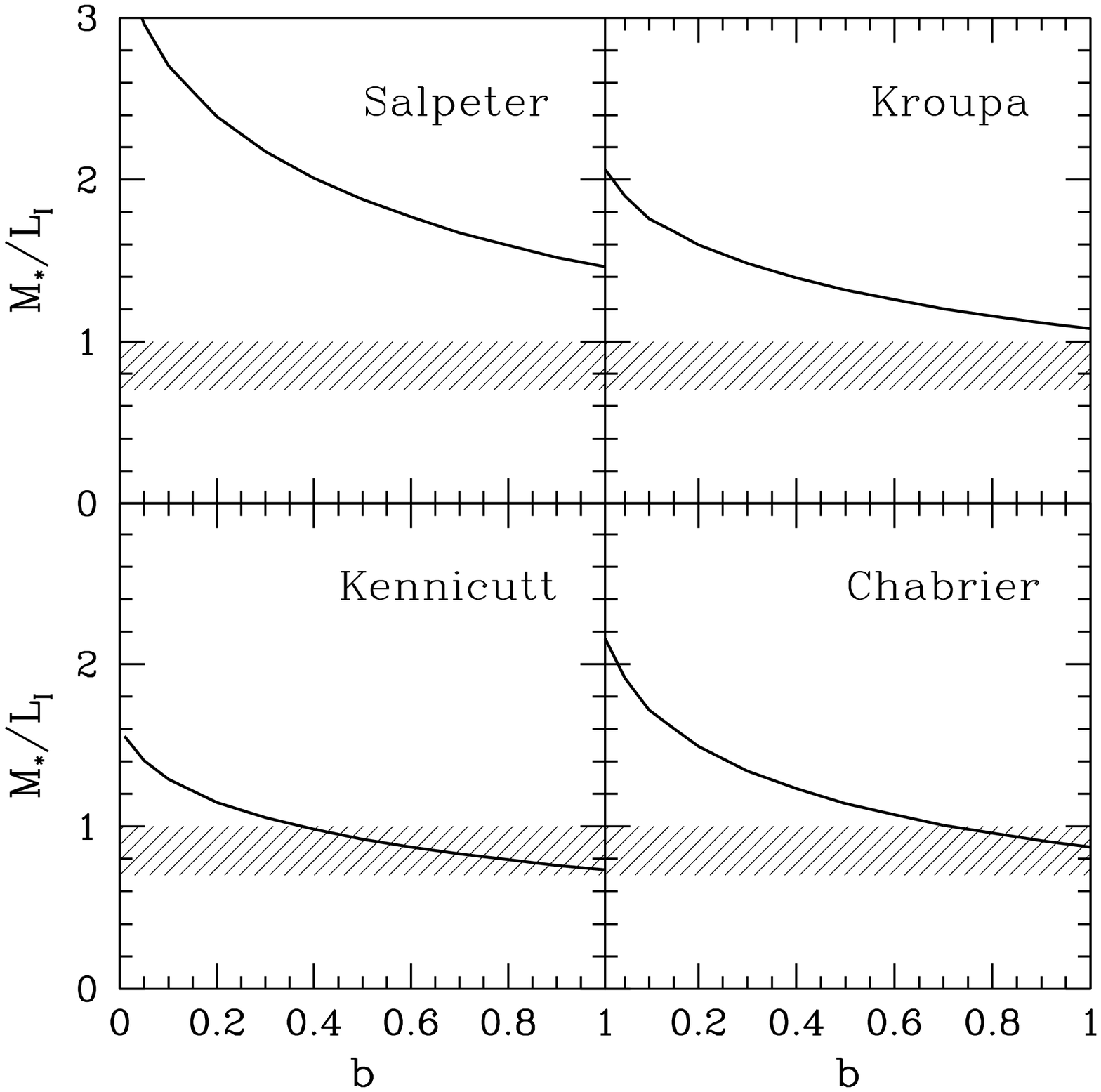,width=6truecm}}
\caption{I--band M/L ratio at varying $b$--parameter of the SFH, for different
IMFs. The red shaded area marks the range M$_*$/L$_I$=0.7--1
observed for Sbc--Sc discs ($b$=0.8--1).}
\label{fig:b-models}}
\end{figure}

\noindent
Cosmological simulations of disc galaxy formation show good agreement
with the observed Tully-Fisher relation,
provided the mass--to--light ratio of the stellar component is 
as low as {\mbox{M$_*$/L$_I$=0.7--1}}; 
a low M$_*$/L is as well 
derived when locating onto the Tully--Fisher relation
{\it real} disc galaxies of known stellar mass, such as the Milky Way or 
NGC 2841 (Fig.~\ref{fig:tully}; Sommer-Larsen et~al.\ 2003; 
Portinari et~al.\ 2004a, hereinafter PST).
Several other arguments 
support a low M$_*$/L in spiral galaxies:

Based on bar instability arguments, Efstathiou et~al.\ (1982) suggest an 
upper limit of {\mbox{M/L$_B \leq 1.5~h$}} for discs, i.e.\ M/L$_B \leq 1$ 
for $h$=0.7
($h$ indicates the Hubble constant H$_0$ in units of
100~km~sec$^{-1}$~Mpc$^{-1}$).

The stellar M$_*$/L 
is also related to the ``maximality'' of discs, i.e.\ to whether 
they dominate or not the dynamics and rotation curves in the inner galactic 
regions.
For his favoured sub--maximal disc model, Bottema (2002) finds
M$_*$/L$_I \sim 0.82$; and even assuming maximal stellar discs, 
lower M$_*$/L ratios are required than those predicted by
the Salpeter Initial Mass Function (IMF)\footnote{Criticism of the Salpeter IMF
is quite unappropriate in a conference held in honour of Ed Salpeter himself. 
Let me thus underline, that criticized in this paper is not
the original result by Salpeter (1955), who derived the IMF slope
between [0.4--10]~$M_{\odot}$; 
but rather what has become in literature 
the default
meaning of ``Salpeter IMF'': a power law with Salpeter slope, extending
over [0.1--100]~$M_{\odot}$.\label{foot:SalpIMF}}
(Bell \& de Jong 2001). 

Finally, recent dynamical studies of individual 
galaxies yield {\mbox{M$_*$/L$\sim$1}} in B, V, I for the Sc NGC 4414 
(Vallejo et~al.\ 2002) and M$_*$/L$_I$=1.1 for the disc of the Sab 2237+0305
(Huchra's lens, Trott \& Webster 2002).

\smallskip \noindent
The M$_*$/L ratio of the stellar component of a galaxy, including both living 
stars and remnants, depends on the stellar IMF and on
the star formation history (SFH) of the system. 

\noindent
{\bf IMF.}~~
Ample observational evidence in the Solar Neighbourhood, in globular and open 
clusters, and in the Galactic bulge show that the IMF presents a bend-over 
below $\sim 1 M_{\odot}$, and is ``bottom--light'' with respect 
to a single--slope Salpeter IMF 
(see the reviews by Scalo, Chabrier, Zoccali and De Marchi in this conference).
A bend--over is expected as well from theory (see 
Session~III in this conference).

In this paper, we consider 
the following IMFs:
the ``{\bf Salpeter}'' IMF (in the sense of footnote~\ref{foot:SalpIMF});
the {\bf Kroupa} (1998) IMF, derived for field stars in the Solar Vicinity;
the {\bf Kennicutt} IMF, derived from the global 
properties of spiral galaxies (Kennicutt et~al.\ 1994);
the {\bf Chabrier} (2001, 2002) IMF, derived from observations of local
low mass stars and brown dwarfs. (Further cases are discussed in PST.)
With respect to Salpeter, the other IMFs are ``bottom--light''; at the high 
mass end, slopes range between the Salpeter ($x=1.35$) and the Scalo 
one ($x=1.7$).

\smallskip 
\noindent
{\bf SFH.}~~
The sequence of Hubble spiral types is a sequence of different SFHs 
in the discs, traced by the birthrate parameter
%
%
{\mbox{$b = SFR/<SFR>$}} i.e.\
the ratio between the present and the past average star formation rate (SFR). 
The observational Tully--Fisher relation in Fig.~\ref{fig:tully}, 
indicating M$_*$/L$_I$=0.7--1, refers to Sbc--Sc spirals whose ``typical'' 
SFH corresponds to $b$=0.8--1 (Kennicutt et~al.\ 1994;
Sommer--Larsen et~al.\ 2003).


\smallskip \noindent
In PST we computed chemo--photometric models for disc galaxies 
predicting the M$_*$/L$_I$ ratio for different IMFs, as a function
of $b$. Fig.~\ref{fig:b-models} shows that, 
while the Salpeter IMF yields far too high M$_*$/L, the other ``bottom--light''
IMFs do yield the observed {\mbox{M$_*$/L$_I$=0.7--1}} for late--type spirals 
($b=0.8-1$), which agrees very well with our present understanding of the shape
of the IMF at the low--mass end.

\begin{figure}[ht]
\sidebyside
{\centerline{ \psfig{file=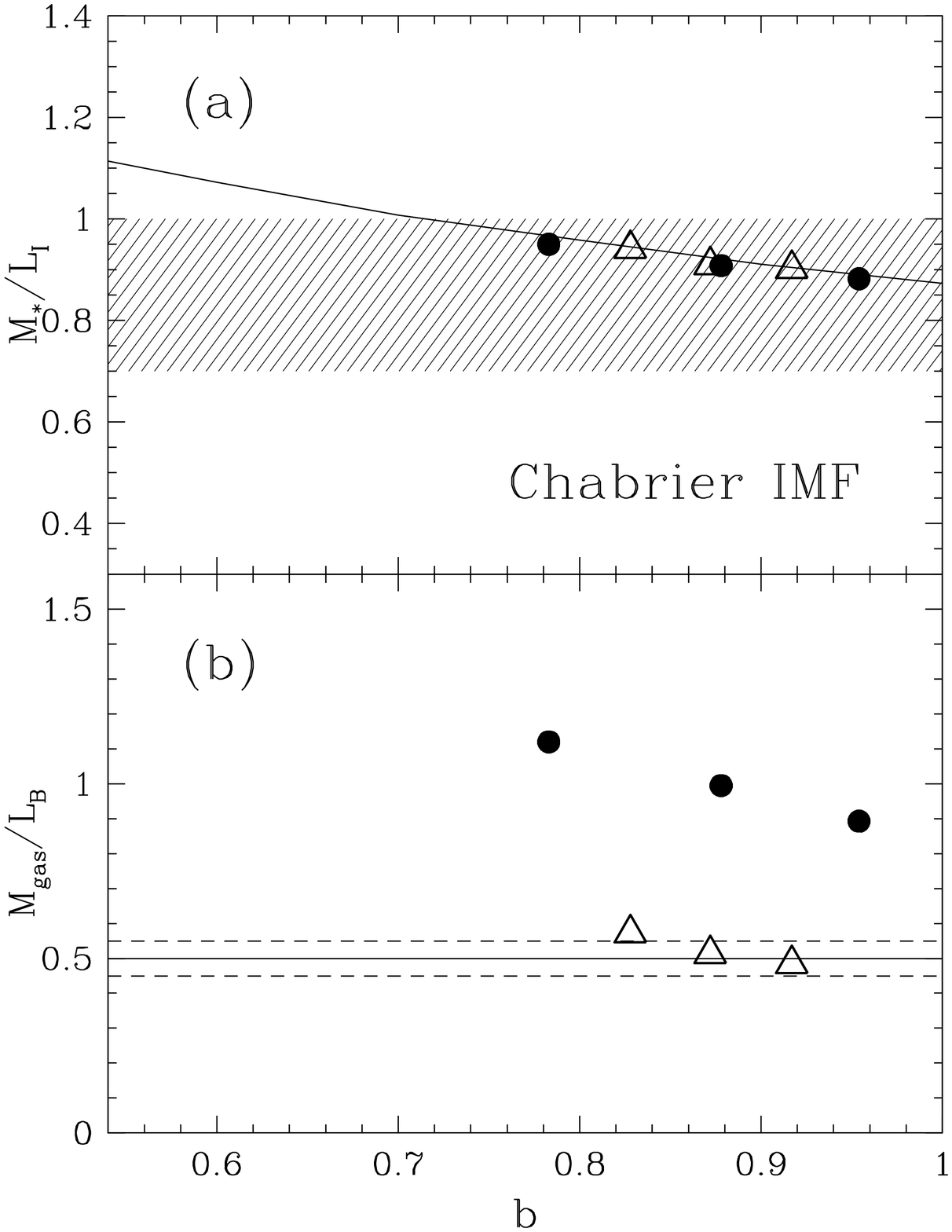,width=5.25truecm} }
\caption{{\bf (a)} 
M$_*$/L$_I$ ratio for 
models with the Chabrier IMF. 
{\bf (b)} Model gas fractions compared to the observed range (dashed lines).
{\em Dots}: models with IMF mass range {\mbox{[0.1--100]~$M_{\odot}$}};
{\em triangles}: models with IMF upper mass limit tuned at
32--33~$M_{\odot}$
to match the observed gas fraction; 
note that 
the effect on M$_*$/L is negligible.}
\label{fig:models-chab}}
{\centerline{\psfig{file=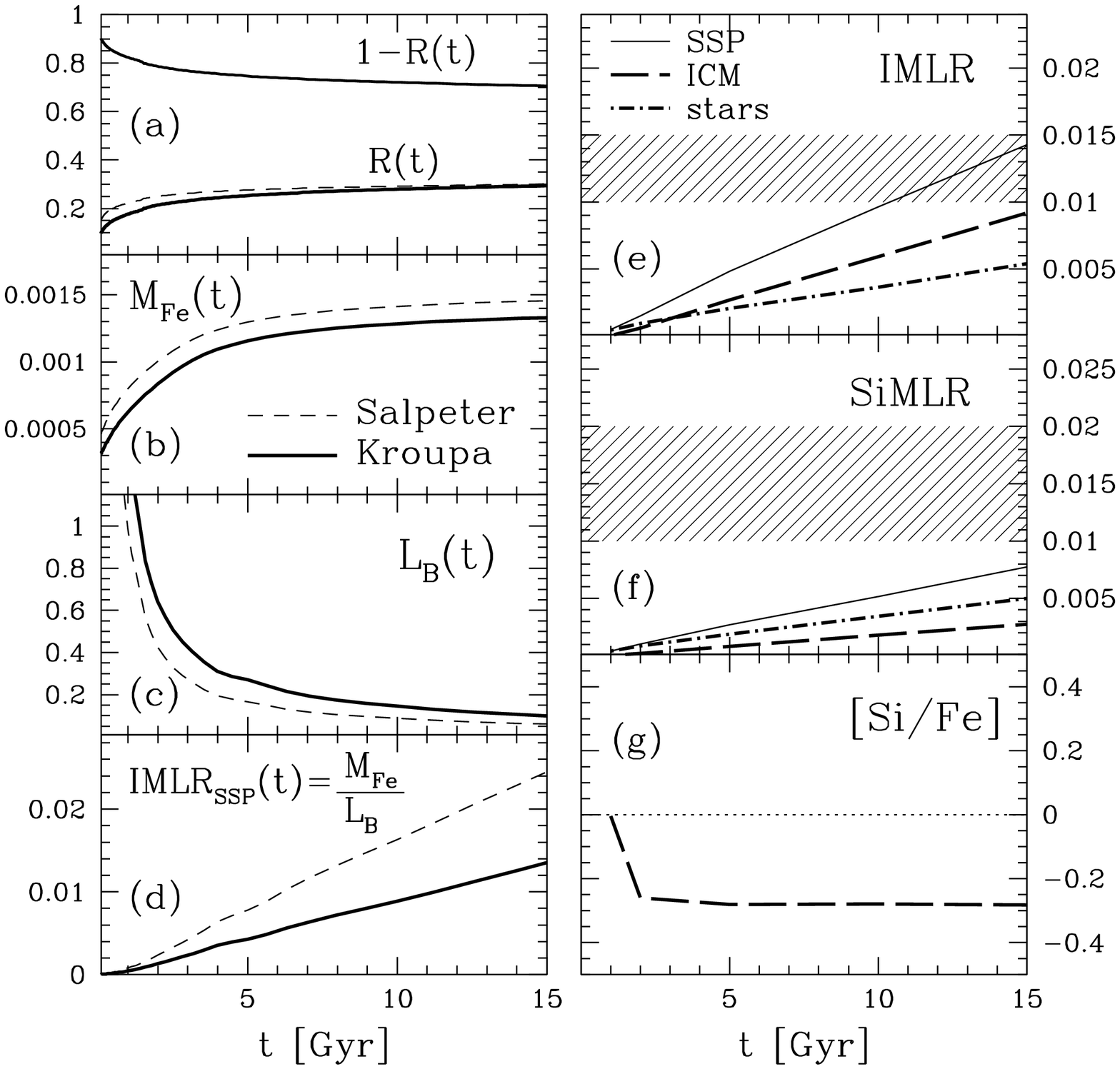,width=7truecm}}
\caption{
{\bf (a--d)} Locked--up fraction, 
iron production, luminosity evolution and IMLR$_{SSP}$ for Salpeter 
and Kroupa IMFs.
{\bf (e--f)} Kroupa IMF: partition of metals 
into stars and ICM, 
compared to observations (shaded area, Finoguenov et~al.\ 2000).
{\bf (g)} Predicted [Si/Fe] in the ICM with the Kroupa IMF.}
\label{fig:kroupaICM}}
\end{figure}

As to the implications for chemical evolution, some ``bottom--light'' IMFs 
(e.g.\ Kennicutt and Chabrier) are too efficient in metal 
production, as is evident from the gas fractions predicted by the models, far
larger than observed 
(Fig.~\ref{fig:models-chab}b): metal enrichment
is so efficient that the models reach the typical metallicities of spirals 
without much gas processing.

This excessive metal production 
is readily understood since the enrichment efficiency of a stellar 
population, or its ``net yield'': 
%
\begin{equation}
\label{eq:yield}
y \, = \, \frac{1}{1-R} \int_{M_i}^{M_s} p_Z(M) \, \Phi(M) \, dM
\end{equation}
%
is inversely proportional to the mass fraction 1--$R$ that remains forever 
locked in low--mass stars and remnants: for bottom--light IMFs 
the locked--up fraction tends to be small.
The yield can be reduced by reducing the number of the massive stars 
responsible for the bulk of the metal production, i.e.\ by tuning the upper 
mass limit (triangles in Fig.~\ref{fig:models-chab}) or by a steep slope 
at the high-mass end: for example, the Kroupa IMF 
is bottom-light but it does not overproduce metals due to the steep Scalo 
slope above $M=1 \, M_{\odot}$ (see PST). 
A steep slope 
for the integrated field stars IMF is expected, in fact, if stars form
in clusters of finite size from an intrinsically shallower IMF 
(Kroupa \& Weidner 2003). 

Alternatively, we need to invoke substantial outflows of metals 
from disc galaxies into the intergalactic medium, to reconcile the high 
enrichment efficiency with the observed metallicities and low gas fractions. 
This behaviour would be reminiscent of that of
elliptical galaxies, responsible for the enrichment of the hot gas in clusters
of galaxies.

\section{The IMF in clusters of galaxies}

\noindent
Can a ``standard'' Solar Neighbourhood (SN) IMF 
account for the observed level of metal enrichment in clusters of galaxies? 
We can qualitatively address this question by 
comparing the respective ``effective'' yield $y_{eff}$, obtained as the ratio 
between the metals globally contained in the system 
and the mass in living stars and remnants; this is the observational 
counterpart of the theoretical yield $y$ in Eq.~\ref{eq:yield}:
\begin{small}
\[ y_{eff, SN} = \frac{Z_* \times M_* + Z_{gas} \times M_{gas}}{M_*}
 \sim \frac{Z_{\odot} \times M_* + Z_{\odot} \times (0.2 \, M_*)}{M_*} 
= 1.2 \, Z_{\odot} \]
%
%
\[ y_{eff,~cl} = \frac{Z_* \times M_* + Z_{ICM} \times M_{ICM}}{M_*}
\sim \frac{Z_{\odot} \times M_* + 0.3 \, Z_{\odot} \times (5-10 \, M_*)}{M_*} 
= 2.5-4 \, Z_{\odot} \]
\end{small}
\noindent
The estimated metal enrichment efficiency in clusters 
is thus about 3 times larger than in the SN, 
and the chemical evolution of clusters 
is often modelled with non--standard IMFs (Portinari et~al.\ 2004b, hereinafter
PMCS; and references therein).
On the other hand, 
the following arguments are often given in favour of a universal
standard IMF:
the Iron Mass--to--Light Ratio
(IMLR) in clusters agrees with the predictions of the Salpeter IMF,
and the observed [$\alpha$/Fe] ratios in the ICM are compatible with those
in the SN
(Renzini et~al.\ 1993; Renzini 1997, 2004; Ishimaru \& Arimoto 1997; 
Wyse 1997). 


In PMCS we demonstrated that a ``standard'' IMF
(e.g.\ the Kroupa IMF) which reproduces the chemical properties of the SN,
is unable to enrich the ICM to the observed levels.
For a Single Stellar Population (SSP) 
we computed the rate of supernov\ae\ of type II and type Ia, 
and the corresponding production of metals in time (Fig.~\ref{fig:kroupaICM}b);
the ratio between these and the evolving luminosity 
(Fig.~\ref{fig:kroupaICM}c) 
gives the global IMLR$_{SSP}$, SiMLR$_{SSP}$ etc.\ relevant to the chosen IMF
(Fig.~\ref{fig:kroupaICM}d).
Not all of the metals produced contribute to the ICM enrichment: 
a non--negligible fraction must be locked-up by
subsequent stellar generations to build up the observed stellar metallicities
$Z_*$ of cluster galaxies. 
The amount of metals locked in the stellar component must be
$M_{Z,*} = Z_* \times (1-R)$, where $1-R$ is the locked--up fraction
consistent with the adopted IMF (Fig.~\ref{fig:kroupaICM}a). 
Once the metals produced are properly partitioned between the stars 
and the ICM, it is evident that a standard
IMF such as the Kroupa IMF cannot possibly reproduce the
observed IMLR and SiMLR in the ICM (Fig.~\ref{fig:kroupaICM}e,f):
it does not match 
the {\it global} amount of metals observed in the ICM and it predicts
significantly sub--solar [$\alpha$/Fe] ratios, at odds with observations
(Fig.~\ref{fig:kroupaICM}g).
Henceforth, observing solar [$\alpha$/Fe] ratios in the ICM 
{\it per se} does not suffice to conclude that the same IMF is at play 
in both environments.

In Fig.~\ref{fig:kroupaICM}d we also compare the global IMLR$_{SSP}$ 
for the Kroupa and the Salpeter IMF. The latter is about twice more efficient
in metal production, and is known to be too efficienct to reproduce the
SN (PST, PMCS and references therein; 
Romano, this conference); henceforth, though matching the IMLR in the ICM, 
it is not the same IMF as in the Milky Way. Besides, the Salpeter IMF (in the
sense of footnote~\ref{foot:SalpIMF}) fails at reproducing the observed
$\alpha$MLR in the ICM, also predicting significantly subsolar [$\alpha$/Fe] 
ratios in the ICM (Matteucci \& Vettolani 1988; Renzini et~al.\ 1993;
Pipino et~al.\ 2002; PMCS).

\section{Conclusions}

\noindent
A ``standard'' IMF suited to model the chemical evolution of the Solar 
Neighbourhood cannot account for the observed metal enrichment in clusters: 
either the
IMF differs between the two environments, or the local IMF has a much
higher yield than usually assumed. The latter option is in line with
some of the ``bottom--light'' IMFs advocated in \S\ref{sect:discIMF}
to reproduce low disc M$_*$/L ratios. In this case, disc galaxies must disperse
much of the metals they produce into the intergalactic medium, just like
early type galaxies in clusters.
However, substantial outflows would challenge our understanding of disc 
galaxy formation: disc star formation proceeds at a smooth,
non burst--like pace and the observed ``fountains'' and ``chimneys'' 
do not have enough energy to escape the galactic potential; winds are far less
plausible than from spheroids. Moreover, 
strong ongoing stellar feedback and outflows could significantly hamper the 
dynamical formation of galactic discs from the cool--out of halo gas.

The alternative scenario is a variable IMF with a higher yield in clusters 
than in disc galaxies.
The IMF may vary after Jeans--mass dependence on redshift, and
its variation should be more
significant than expected from the increasing temperature of the cosmic 
background (e.g.\ Chabrier, this conference; Finoguenov et~al.\ 2003; 
Moretti et~al.\ 2003 and references therein);
or, the IMF may be a universal function within star clusters, but
generating statistically more high--mass stars in larger star clusters and 
in regimes of intense star formation like in massive ellipticals
(Kroupa \& Weidner 2003).



\begin{chapthebibliography}{}


\bibitem[]{}
Bell E.F.\ \& de Jong R.S., 2001, ApJ 550, 212

\bibitem[]{} 
Bottema R., 2002, A\&A 388, 809

\bibitem[]{}
Chabrier G., 2001, ApJ 554, 1274

\bibitem[]{}
Chabrier G., 2002, ApJ 567, 304

\bibitem[]{} 
Dale D.A., Giovanelli R., Haynes M.P., et~al.,
1999, AJ 118, 1489

\bibitem[]{} 
Efstathiou G., Lake G.\ \& Negroponte J., 1982, MNRAS 199, 1069


\bibitem[]{} 
Finoguenov A., David L.P.\ \& Ponman T.J., 2000, ApJ 544, 188

\bibitem[]{} 
Finoguenov A., Burkert A.\ \& B\"ohringer H., 2003, ApJ 564, 136



\bibitem[]{} 
Ishimaru Y., Arimoto N., 1997, PASJ 49, 1


\bibitem[]{} 
Kennicutt R.C., Tamblyn P.\ \& Congdon C.W., 1994, ApJ 435, 22 (KTC94)

\bibitem[]{} 
Kroupa P., 1998, 
ASP Conf.\ Series vol.~134, p.~483

\bibitem[]{} 
Kroupa P.\ \& Weidner C., 2003, ApJ 598, 1076


\bibitem[]{} 
Matteucci F., Vettolani P., 1988, A\&A 202, 21

\bibitem[]{} 
Moretti A., Portinari L.\ \& Chiosi C., 2003, A\&A 408, 431




\bibitem[]{} 
Pipino A., Matteucci F., Borgani S.\ \& Biviano A., 2002, NewA 7, 227



\bibitem[]{} 
Portinari L., Sommer--Larsen J.\ \& Tantalo R., 2004a, 
{\mbox{MNRAS}} 347, 691 (PST)

\bibitem[]{} 
Portinari L., Moretti A., Chiosi C.\ \& Sommer--Larsen J., 2004b, 
ApJ 604, 579 (PMCS)

\bibitem[]{} 
Renzini A., 1997, ApJ 488, 35

\bibitem[]{} 
Renzini A., 2004, in Clusters of galaxies: probes of cosmological structure
and galaxy evolution, ed.\ J.~Mulchaey et~al.\ 
(Cambridge University Press), p.~261

\bibitem[]{} 
Renzini A., Ciotti L., D'Ercole A.\ \& Pellegrini S., 1993, ApJ 419, 52


\bibitem[]{} 
Salpeter E.E., 1955, ApJ 121, 161



\bibitem[]{}
Sommer--Larsen J., G\"otz M.\ \& Portinari L., 2003, ApJ 596, 47


\bibitem[]{}
Trott C.M.\ \& Webster R.L., 2002, MNRAS 334, 621

\bibitem[]{}
Vallejo O., Braine J.\ \& Baudry A., 2002, A\&A 387, 429

\bibitem[]{}
Wyse R.F.G., 1997, ApJ 490, L69

\end{chapthebibliography}
\end{document}